\documentclass[english]{article}
\usepackage[utf8]{inputenc}

\usepackage{bm}
\usepackage{xcolor}
\usepackage{amsmath,amssymb}
\usepackage{authblk}
\usepackage{array}
\usepackage{dsfont}
\usepackage{graphicx}
\usepackage{rotating}
\usepackage{mathtools}
\title{Composite operators of stochastic model A}

\author[1]{D.~Davletbaeva}
\author[2,3,4]{M.~Hnati\v{c}}
\author[1]{M. V. Komarova}
\author[3]{T.~Lu\v{c}ivjansk\'y}
\author[2]{L. Mi\v{z}i\v{s}in}
\author[1,2]{M. Yu. Nalimov}

\affil[1]{Department of Theoretical Physics, St. Petersburg University, St. Petersburg, Russia}
\affil[2]{Bogoliubov Laboratory of Theoretical Physics, Joint Institute for Nuclear Research, Dubna, Russia}
\affil[3]{Faculty of Science, \v{S}af\'arik University, Ko\v{s}ice, Slovakia}
\affil[4]{Institute of Experimental Physics SAS, Ko\v{s}ice, Slovakia}

\newcommand{\ppsi}{\psi^{+}}
\newcommand{\ppsic}{\psi^{+'}}
\newcommand{\psic}{\psi^{'}}

\newcommand{\ff}{\varphi}
\newcommand{\ffi}{\varphi'}
\newcommand{\action}{\mathcal{S}}
\newcommand{\vf}[1]{F_{#1}}
\newcommand{\veps}{\varepsilon}
\newcommand{\DRG}{\mathcal{D}_{RG}}

\def\unitM{\mathds{1}}
\allowdisplaybreaks

\begin{document}

\maketitle
\begin{abstract}
By means of the field-theoretic renormalization group, we study
 the damping of the viscosity coefficient near the superfluid phase transition.
 We utilize the fact that in the infrared region, the complex model used to describe the phase transition belongs to the same universality class as the well-known stochastic model~A. This allows us a determination of the critical behavior of viscosity using composite operators for model~A. Our analysis is based on the $\veps$-expansion near the upper critical dimension $d_c =4$ of model~A.
 The critical exponent of viscosity is then calculated from the critical dimensions of composite operators of massless two-component model~A. In particular, we present results for critical dimensions of a selected class of composite operators with the canonical dimension $8$ to the leading order.
\end{abstract}

\section{Introduction}

 The most relevant features of the system that affect its critical behavior are gross
 properties such as symmetries of the order parameter and fields, the tensor character of the order parameter, the number of components, etc. These to a large extent determine to what kind of universality class a given model belongs \cite{Vasilev,Zinn2002,Tauber}. According to the classical monographs \cite{Hohenberg1977,Folk2006}  superfluid critical dynamics  should be properly captured  by stochastic dynamical models E or F. Fundamental dynamical variables of these models are
  slowly varying fields that correspond to thermodynamic flows \cite{Hohenberg1977} and corresponding
  theoretical models are conveniently expressed through them. However, for model~E, the renormalization group (RG) equation leads to the two candidates for an infrared (IR) stable fixed point, and the question of which of them actually determines the observed behavior of the bosonic gas remains  open (unclear) \cite{Vasilev,peliti1978}. In addition, neither in model~E nor in the more involved  F~model, it is possible to calculate the critical exponent that determines the damping of coefficient viscosity when approaching the critical point of the corresponding phase transition from normal to the superfluid state. 


In the recent papers \cite{Hnatich2013,Danco2016} it has been proposed to analyze the stability of  models E and F with respect to the influences of the additional hydrodynamic velocity field, and 
due to density wave perturbations, respectively. It has turned out that both models are sensitive to the inclusion of hydrodynamic modes, which significantly affect the critical behavior and, in particular, values of the critical exponents. Further, taking into account the compressibility of the medium led to a paradoxical conclusion \cite{Zhavoronkov2019}: models~E and F of critical dynamics are unstable with respect to such perturbations, and in the IR region effectively reduce to model~A \cite{Vasilev,Tauber,Hohenberg1977}, which exhibits a single IR stable fixed point. On the other hand, phenomenological considerations are confirmed from a microscopic point of view \cite{Honkonen2019,Honkonen2022}, where a careful analysis of the Bose gas was carried out based on the formalism of
time-dependent Green functions at finite temperature.
This formalism provides additional support in favor of the aforementioned results.


The main conclusion of the corresponding analysis \cite{Zhavoronkov2019} is that model~A also correctly describes dynamical behavior in the vicinity of the lambda point. 
Moreover, the original effective dynamical action determines critical dimensions of IR irrelevant fields (e.g, velocity field 
${\bm v}$, or field $m$ that corresponds the density and temperature fluctuations) or possible monomials constructed from them, in terms of dimensions of composite operators of the two-component model~A. In other words, once the most important IR composite operators of the main fields are determined, it is possible to analyze the damping of the viscosity coefficient using the composite operators of model~A. 

By employing sophisticated RG techniques on model~A \cite{Vasilev,Hohenberg1977} it is possible to calculate the critical exponent of viscosity in terms of the critical dimensions of the composite operators with canonical dimension $8$. Composite operators with lower canonical dimensions $2, 4$, and $6$ are mixed with them, and missing powers are necessarily compensated with the powers of a parameter (chemical potential or momenta) responsible for the phase transition in the original model. Our main aim here is to calculate critical dimensions of operators of canonical dimension $8$ to the leading order in $\veps = 4 - d$ expansion.  Their determination  allows us to obtain physically relevant information about the damping of viscosity in the vicinity of the phase transition from the normal state to the superfluid state.

\section{Stochastic Model A}

In the standard classification  of specific dynamical models \cite{Hohenberg1977}, the model~A can be regarded as the simplest model. For a long time, it was mainly used to describe the critical behavior of an anisotropic ferromagnet or antiferromagnet. 
 The action functional for the generalized model~A takes the abbreviated form  \cite{Vasilev}
\begin{equation}
\action (\ffi, \ff )= \alpha_0 \ffi^2 + \ffi \left(- \frac{\partial}{\partial t} \ff +\alpha_0 \nabla^2 \ff - \alpha_0 \tau_0 \ff - \frac{\alpha_0 g_0}{6}\ff^3 \right),
\label{eq:action_A}
\end{equation}
where fields $\ff$ ($\ffi$) is $n$-component order parameter (response) field,
$\tau_0 \propto (T -T_c)$ is (unrenormalized) deviation from critical temperature, $\alpha_0>0$ is kinematic (diffusion) coefficient, and the $g_0$ is a coupling constant of the theory. 

Correspondence between the original effective dynamical action \cite{Zhavoronkov2019,Honkonen2019} and  generalized model~A is straightforward and reads
\begin{equation}
  \ppsic = \ffi_1,  \psic = \ffi_2,  \psi = \ff_1,  \ppsi = \ff_2.
  \label{eq:correspondence}
\end{equation}
Therefore, the calculation of the critical behavior of viscosity is effectively described by 
the two-component model~A. We will, however, continue to work with the generalized model~A. 
The original effective action \cite{Zhavoronkov2019,Honkonen2019} does not contain the "mass" term proportional to $\tau$ (deviation from critical temperature $\tau = T-T_c$). Hence we also set
$\tau = 0$ in the generalized model \eqref{eq:action_A}.

\begin{figure}[!ht]
\begin{center}
\includegraphics[width=5cm]{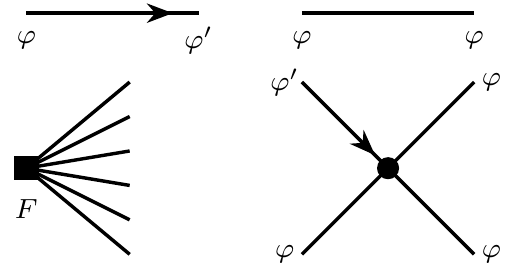}
\end{center}
\caption{Graphical representation of Feynman rules for model~A with general notation for a composite operator $F$.}
\label{fig:Feynman_rules}
\end{figure}
  
The calculation follows the standard procedure of quantum field theory \cite{Vasilev,Zinn2002}.  Feynman diagrammatic rules and all related diagrammatic elements can be obtained in a
straightforward manner from the action functional \eqref{eq:action_A}. 
Propagators can be derived from the Gaussian part of the action 
and, in the frequency-momentum representation, they read
\begin{eqnarray}
  \langle \ff \ffi \rangle = \frac{1}{- i \omega + \alpha k^2}, \quad
  \langle \ff \ff \rangle = \frac{2\alpha}{|-i\omega + \alpha k^2|^2}.
  \label{eq:propagators}
\end{eqnarray}
The interaction part consists of a single vertex $\ffi \ff^3 $. All graphic elements of the perturbation theory are depicted schematically in  Fig.~\ref{fig:Feynman_rules}.

\begin{table}[!ht]
\begin{center}
\setlength{\extrarowheight}{2pt}
\begin{tabular}{||c||c|c|c|c|c|c|c|c||}
\hline 
$Q$ & $ \partial_{x}$ & $\partial_{t}$  & $\ff$ & $\ffi$ & $\alpha$ & $g_0$  \\[1mm]
\hline
$d_{p} [Q] $ & $1$ & $0$ & $\frac{d}{2} - 1$ & $\frac{d}{2} + 1$ & $-2$ & $4-d$ \\ [1mm]
\hline
$d_{\omega} [Q]$ & $0$  & $1$ & $0$ & $0$ & $1$ & $0$  \\ [1mm]
\hline
$d [Q] $ & $1$ & $2$ & $\frac{d}{2} - 1$ & $\frac{d}{2} + 1$ & $0$ & $4-d$ \\ [1mm]
\hline
\hline
\end{tabular}
\end{center}
\caption{Canonical dimensions of fields and parameters for the model~A. }
\label{tab:can_dim_par}
\end{table}

The starting point of the field-theoretical RG approach is dimensional analysis. As a rule, dynamical models of type \eqref{eq:action_A} exhibit two independent scales \cite{Tauber,Vasilev}. This means that to each quantity $Q$, two independent canonical dimensions can be assigned, i.e., the momentum dimension $d_p [Q]$ and the frequency dimension $d_{\omega} [Q]$, respectively.
 They are determined from the requirement that each term in the action functional \eqref{eq:action_A} has to be dimensionless \cite{Vasilev,Zinn2002}. We employ the standard normalization conditions (for canonical dimensions of momentum $p$ and frequency $\omega$) in the following form 
 \begin{align}
  d_p [p] & = - d_p [x] = 1, & d_p [\omega] & =  d_p [t] = 0, 
  \label{eq:canon1}  
  \\
  d_{\omega} [p] & = d_{\omega} [x] = 0, & d_{\omega} [\omega] & =  - d_{\omega} [t] =1. 
  \label{eq:canon2}
\end{align}
The total canonical dimension of a quantity $Q$ is given by a formula
\begin{equation}
  d [Q] = d_p [Q] + 2 d_{\omega} [Q],
  \label{eq:can_dim}
\end{equation}
and all canonical dimensions for model~A are summarized in Tab.~\ref{tab:can_dim_par}. 
From which we also infer that is that the upper critical dimension of the model is $d_c = 4$, for
which the charge $g_0$ becomes dimensionless.
This motivates us to introduce the formally small parameter $\veps$ of the theory \cite{Vasilev,Zinn2002} as a deviation 
\begin{equation}
  \veps = d_c -d = 4 - d .
\end{equation}  
 We perform the entire RG analysis to the leading order of the perturbation theory ($\sim g$)
 employing the minimal subtraction (MS) scheme \cite{Vasilev,Zinn2002}. 

Since more information about the renormalization of generalized model A can be found in the literature \cite{Vasilev,Zinn2002,Hohenberg1977}, we do not discuss it here. 
Rather, our main aim  
is to analyze composite operators and their construction, and perform related renormalization procedure. 

\section{Extended model}

In the IR region, the original model with included hydrodynamic fluctuations \cite{Zhavoronkov2019,Honkonen2019} reduces to the generalized model A \eqref{eq:action_A}. 
In this reduction, the terms containing the viscosity coefficient are not a part of model A. 
However, they can be interpreted as an extension of the model by certain composite operators. From the point of view of quantum field theory, the composite operator $F$ is an arbitrary local scalar quantity (monomials) constructed from fields and their derivatives. Let us call the model with included composite operators extended.
 For such a model, the generating functional takes the form  
\begin{equation}
  G(A, A', a) = c \int D\ffi D\ff \, \exp\left[ \action(\ff,\ffi) + aF(\ff,\ffi) + A \ff 
  + A' \ffi\right],
\end{equation}
where $a$ is a set of sources and the linear form $a F(\ff, \ffi)$ is a convenient abbreviation for the expression $ \sum \int\, dx dt \ a_i (t, x) F_i (t, x; \ff, \ffi)$. Since all the main theorems of the renormalization are formulated for an arbitrary local interaction, they remain valid even for the extended model with interaction part including the composite operators $(\alpha g/6 \ffi \ff^3 + a F)$ \cite{Vasilev}. 


As has been mentioned previously, in the renormalization analysis the first step is canonical dimension analysis. Here the original effective dynamical model \cite{Zhavoronkov2019,Honkonen2019} comes into consideration, for which the maximum value of the total canonical dimension for the term with viscosity is $d_{max} [F] = 8$. However, during the renormalization of composite operators, other operators with a lower canonical dimension also contribute to the corresponding counterterms. And thus it is necessary to take into account operators with a lower dimension. On the other hand, a monomial has to be a scalar quantity, which is valid only for monomials with even dimensions. This reduces composite operators to monomials with dimension $d = \{ 2, 4, 6, 8\}$. 
Another simplification comes from the fact that model A is studied at the critical
point $\tau = 0$. This has the consequence that operators with different canonical dimensions do not mix with each other in the renormalization process. In other words, composite operators are divided into sets according to their total canonical dimension, 
and operators in a given set do not give contributions to operators in a different set during the renormalization procedure. In this article, we focus on the RG analysis of the operators with canonical dimension $8$.

Each composite operator is a scalar monomial, with two independent canonical dimensions and its total canonical dimension is determined using Eq. \eqref{eq:can_dim}, which corresponds to the sum of the dimension of the fields and the derivations defined in Tab. \ref{tab:can_dim_par}. The canonical dimension of the source $a_i$ can be determined from the condition that the action $\action$ is a dimensionless quantity and thus we get
\begin{equation}
 d [a] + d [F] + d [x] + d [t] = d [\action] \equiv 0 .
\end{equation}
Replacing $d \rightarrow \Delta$ yields a similar relation between critical dimensions.
 In this way, we can determine the critical dimensions of sources $\Delta_a$ once
  we know the critical dimension parameters and composite operators $\Delta_F$. 


The construction of all possible monomials of fields and their derivatives, whose canonical dimension is $8$ leads to $74$ different scalar monomials. Let us note that they contain a single derivative, which acts only on one of the fields $\ffi$ or $\ff$, respectively. Composite operators containing multiple derivatives can be expressed through a suitable combination of operators with a single derivative. It was necessary to take into account that the fields are $n$-component and therefore we need to distinguish the product between fields in the monomials. In actual calculations, we work with a general $n$-component field and only at the end of the calculation we set $n=2$.

%
%
 The composite operators can be divided into different sets. One of them is formed by composite operators, collectively denoted by $O$, containing exactly two fields (e.g., $\ffi \partial_t \ffi$, $\ffi \partial_t \partial^2 \ff$, $\ff \partial^6 \ff$, and so on). Operators $O$ display two important properties in the leading order of perturbation theory. First, operators $O$ form a closed set with the canonical dimension $8$, which means that composite operators with different numbers of fields do not mix with these $O$ operators. Second, the information about the renormalization of $O$ operators can be obtained from the renormalization of operators with lower canonical dimensions, i.e. those with canonical dimensions equal $6$, $4$, or $2$.
 These two features can be proved by the direct inspection of Feynman diagrams and lead to the fact that there is, in fact, no need for the renormalization  of $O$ operators. In other words, the composite operator $O$ can be excluded from the renormalization procedure because its renormalization is extracted from an
 operator with a lower canonical dimension. 
%
%

 As a result of these considerations, the final number of necessary composite operators is  reduced to a closed set of $41$ operators, which we summarized in the following way

\begin{align}
\vf{1} & = (\ffi \ffi)(\ff \ff),  & \vf{2} & = (\ffi \ff)(\ffi \ff),  & \vf{3} & = (\ff \partial^2 \ffi) (\ff \ff), \nonumber \\ 
\vf{4} & = (\ff \partial_i \ffi) ( \ff \partial_i \ff), & \vf{5} & = (\partial_i \ffi  \partial_i \ff) ( \ff  \ff), & \vf{6} & = (\ffi \partial^2 \ff) (\ff \ff), \nonumber \\ 
\vf{7} & = (\ffi \ff) (\ff \partial^2 \ff),  & \vf{8} & = (\ffi \ff) (\partial_i \ff)^2, &
\vf{9} & = ( \ffi \partial_i \ff) (\ff \partial_i \ff), \nonumber \\ 
\vf{10} & = (\ffi \partial_t \ff) (\ff \ff), & \vf{11} & = (\ffi \ff) (\ff  \partial_t \ff), & \vf{12} & = (\ff \partial_t \ffi) (\ff \ff), \nonumber \\ 
\vf{13} & = (\ffi \ff) (\ff \ff)^2,   & \vf{14}  & = (\ff \partial_t \ff) (\ff \ff)^2,  & \vf{15} & = (\ff \partial_t^2 \ff) (\ff \ff),  \nonumber \\ 
\vf{16} & = (\partial_t \ff \partial_t \ff) (\ff \ff),  & \vf{17} & = ( \ff \partial_t \ff) (\ff \partial_t \ff),   & \vf{18} & = (\ff \partial_t \partial^2 \ff) (\ff \ff), 
\nonumber\\ 
\vf{19} & = (\partial_t \ff \partial^2 \ff) (\ff \ff),   & \vf{20} & =  ( \ff \partial^2 \ff) (\ff \partial_t \ff), & \vf{21} & = (\ff \partial_t \ff) (\partial_i \ff \partial_i \ff), \nonumber \\ 
\vf{22} & = (\partial_i \ff \partial_t \ff) ( \ff \partial_i \ff), & \vf{23} & = (\partial \ff \partial_t \partial \ff) (\ff \ff),  & \vf{24} & = ( \ff \partial_t \partial_i \ff) (\ff \partial_i \ff),  \nonumber \\ 
\vf{25} & = (\ff \ff)^4, & \vf{26} & = (\ff \partial^2 \ff) (\ff \ff)^2,  & \vf{27} & =(\partial_i \ff \partial_i \ff) (\ff \ff)^2,  \nonumber \\
\vf{28} & = (\ff \partial_i \ff)^2 (\ff \ff),  & \vf{29} & = (\ff \partial^4 \ff) (\ff \ff),  & \vf{30} & = (\partial^2 \ff \partial^2 \ff) (\ff \ff), \nonumber \\ 
\vf{31} & = ( \ff \partial^2 \ff) (\ff \partial^2 \ff), & \vf{32} & = (\partial_i \ff \partial_i \partial^2 \ff) (\ff \ff), &  \vf{33} & = (\ff \partial_i \partial^2 \ff) (\ff \partial_i \ff), \nonumber \\ 
\vf{34} & = (\partial_i \ff \partial_i \ff) (\ff \partial^2 \ff),   & \vf{35} & = (\partial^2 \ff \partial_i \ff) ( \ff \partial_i \ff), & \vf{36} & =  (\partial_i \ff \partial_i \ff) (\partial_j \ff \partial_j \ff),  \nonumber  \\
\vf{37} & = (\partial_i \partial_j \ff \partial_i \partial_j \ff) (\ff \ff), & \vf{38} & = ( \ff \partial_i \partial_j \ff) (\ff \partial_i \partial_j \ff), &  \vf{39} & = (\partial_i \ff \partial_j \ff) (\ff \partial_i \partial_j \ff), \nonumber \\
\vf{40} & = (\partial_i \partial_j \ff \partial_i  \ff) (\ff \partial_j \ff),  & \vf{41} &  =  (\partial_i \ff \partial_j \ff) (\partial_i \ff \partial_j \ff), 
\label{eq:composite_operators}
\end{align}
%
%
where for simplicity we have denoted time derivative as ${\partial}/{\partial t} \equiv \partial_t$ and spatial derivative as ${\partial}/{\partial {\bf x}_i} \equiv \partial_i$. Summation over the same index is always implied, e.g. $\partial_i \partial_i =\partial^2 \equiv \nabla^2$. The canonical dimensions for composite operators are listed in Tab. \ref{tab:can_dim_f}. The composite operators are chosen in such a way that the derivative is always applied on one of entering fields in a given
composite operator. This to some extent simplifies calculational procedure.

Further, any other operator of canonical dimension $8$ can be expressed by proper linear combination (e.g. $\partial_t [(\ff \ffi)(\ff \ff)] = \vf{10} + 2 \vf{11} + \vf{12}$, $\partial_t (\ff \ff)^3 = 6 \vf{14}$, $ (\ff \ff) \partial^2 (\ff \ffi) = \vf{3} +2 \vf{5} + \vf{6} $, etc.). Although the composite operators \eqref{eq:composite_operators} form a closed set, for some of them their renormalization can be related to other operators with lower canonical dimension, for instance $ \partial^2 [(\ff \ff)^3]_R = [\partial^2 (\ff \ff)^3]_R = [\partial_t (\ff \ff)^3]_R$. This property might be exploited to reduce the number of independent operators, which we have used to provide an independent check  in actual calculations. 
%
%

\begin{table}[!ht]
\begin{center}
\begin{tabular}{||c||c|c|c||}
\hline 
$F_i$ & $F_{1} - F_{9}, F_{29}  - F_{41}$ & $F_{10} - F_{12}, F_{18} - F_{24}$  & $F_{13}, F_{26} - F_{28}$  \\[1mm]
\hline
$d_{p} [F_i] $ & $2d$ & $2d -2$ & $3d -4$  \\ [1mm]
\hline
$d_{\omega} [F_i]$ & $0$ & $1$ & $0$  \\ [1mm]
\hline
$d [F_i] $ & $2d$ & $2d$ & $3d - 4$ \\ [1mm]
\hline
\hline
$F_i$ & $F_{14}$ & $F_{15}-F_{17}$ & $F_{25}$ \\[1mm]
\hline
\hline
$d_{p} [F_i] $ & $3d - 6$ & $2d -4$ & $4d -8$  \\[1mm]
\hline
$d_{\omega} [F_i] $  & $1$ & $2$ & $0$  \\[1mm]
\hline
$d [F_i] $ & $3d - 4$ & $2d$ & $4d - 8$ \\[1mm]
\hline
\end{tabular}
\end{center}
\caption{Canonical dimension of the composite operators $F_i;i=1,\ldots,41$.  }
\label{tab:can_dim_f}
\end{table}
\begin{figure}[!ht]
\begin{center}
\includegraphics[width=0.9\linewidth]{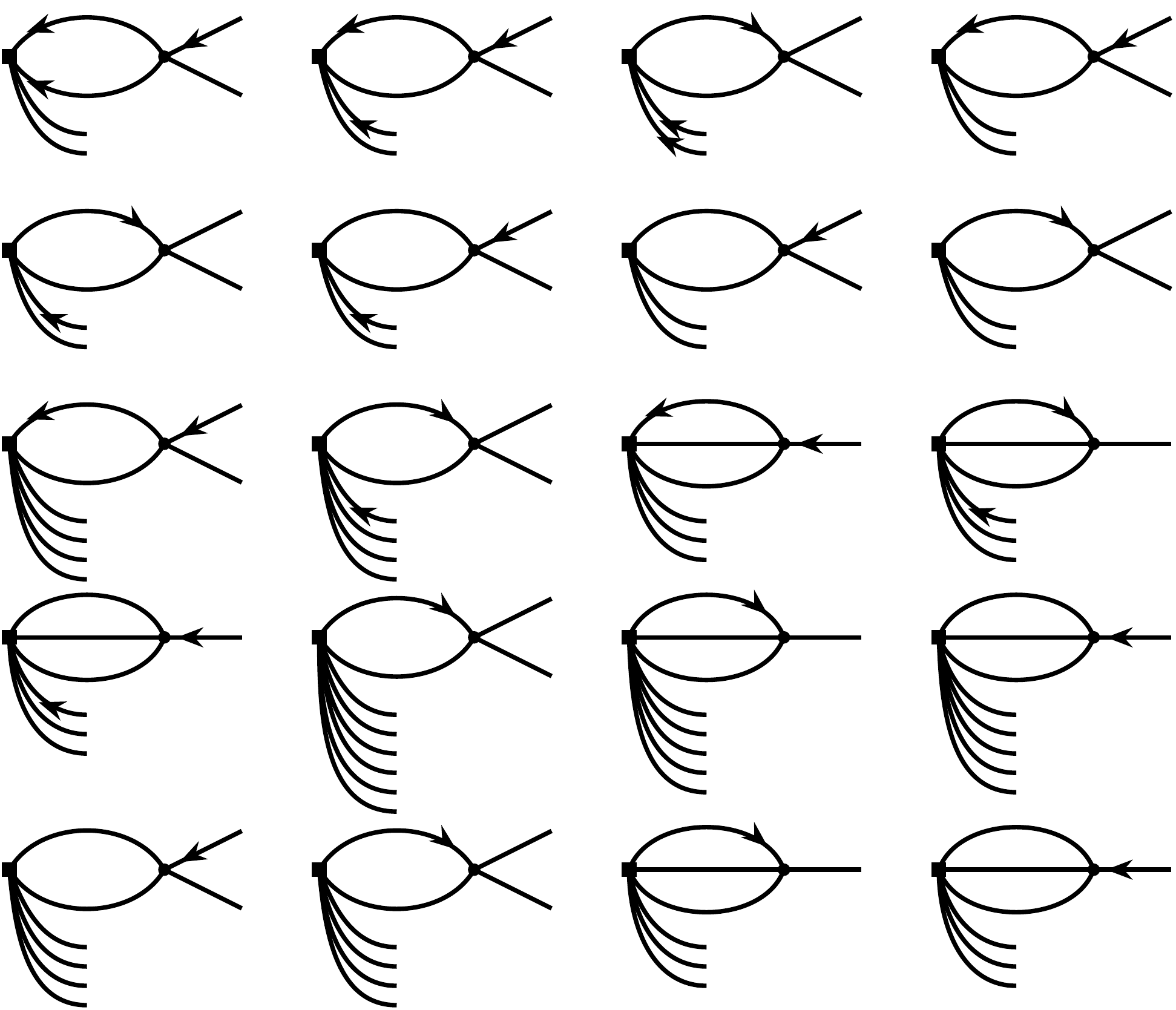} \\
\end{center}
\caption{Feynman diagrams contributing to the composite operator \ref{eq:composite_operators} in the leading order.}
\label{fig:oper_diag}
\end{figure}


The model \eqref{eq:action_A} without the term $a F$ is multiplicatively renormalizable and the expression for multiplicative renormalization through terms of the first order in $a$ can be generalized to the extended model. 
For a closed set of composite operators, the relation between renormalized and unrenormalized takes the general form
\begin{equation}
[F_i (\ffi ,\ff) ]_R = \sum\limits_k \mathcal{Q}_{ik} F_k (\ffi, \ff).
\end{equation} 
The mixing matrix $\mathcal{Q}$ is calculated directly from Feynman diagrams involving composite operators (see Fig.~\ref{fig:oper_diag}). Moreover, the mixing matrix breaks down into three parts in the leading order of the perturbation theory, in which the composite operators mix with each other during RG procedure.
Its elements depend on the renormalization mass $\mu$ and the renormalized parameters 
$\{g, \alpha \}$. For the closed set \eqref{eq:composite_operators}, the mixing matrix can be schematically  written as follows
\begin{equation}
\mathcal{Q} (36 \times 36) = \unitM - \frac{g}{\veps}
\begin{pmatrix}
\mathcal{Q}_{1} (13 \times 13) & 0  &  0 \\
? & \mathcal{Q}_{2} (11 \times 11) & 0 \\
? & ? &  \mathcal{Q}_{3} (17 \times 17) 
\end{pmatrix}
,
\label{eq:qmatrix}
\end{equation}
where $\unitM$ is a unit matrix and $\mathcal{Q}_1, \mathcal{Q}_2, \mathcal{Q}_3$ are block
 matrices with size indicated in brackets (see Appendix). For the  matrix, $\mathcal{Q}$, the rows and columns  correspond to the sequence of the composite operator $\{F_i\}$ from the set \eqref{eq:composite_operators}. The $"0"$ is a notation for the zero blocks in the matrix and the symbol~$"?"$ is part of the  matrix, that does not need to be considered.

Calculation of the diagrams is carried out in the MS scheme \cite{Vasilev,Zinn2002}. The two-loop diagrams (see Fig.~\ref{fig:oper_diag}) do not actually need to be calculated. In fact, they contribute to the matrix \eqref{eq:qmatrix} in the part marked with the symbol $"?"$. The blocks of the mixing matrix $\mathcal{Q}$, 
\begin{itemize}
\item $\mathcal{Q}_1$ - mixing of operators composed of $\ffi, \ff, \partial_t$, and $\partial$,
\item $\mathcal{Q}_2$ - mixing of operators composed of $\ff, \partial_t$, and $\partial$,
\item $\mathcal{Q}_3$ - mixing of composite operators composed of $\ff, \partial$. This corresponds to the static $\ff^4$ model.
\end{itemize}
As has been already stressed these blocks can be analyzed separately, which substantially
simplifies overall analysis.

Further, we continue with the calculation of the critical dimensions for the set of composite operators. The first step is to determine renormalization constants from mixing matrix $\mathcal{Q}$.  The renormalization constants for sources $a_0 = a Z^a$ become a matrix and there
is a relation with the mixing matrix 
\begin{equation}
Z^a_{ij} = \mathcal{Q}_{ij} (Z_{\ff})^{-n_j} (Z_{\ffi})^{-m_j}, \quad
Z_F = (Z^a)^{-1},
\label{eq:rg_constant}
\end{equation}
where $n_j$ and $m_j$ is a number of fields $\ff$, $\ffi$ appearing in monomials $F_j$, respectively. The renormalization constant $Z_F$ is the inverse matrix of $Z^a$. The renormalization constants for fields $Z_{\ff}, Z_{\ffi}$ do not have poles in the MS scheme for the generalized model~A in the leading order, which leads to the relations
\begin{equation}
Z^a  = \mathcal{Q}, \quad Z_F = \mathcal{Q}^{-1}.
\end{equation}
where $\mathcal{Q}^{-1}$ is an inverse mixing matrix.

The information about critical scaling can be obtained from RG equations for composite operators.  They can be derived from the generating functional of renormalized connected Green functions of the fields and composite operators \cite{Vasilev,Zinn2002}.
Using RG method it can be shown that composite operators obey the matrix equation
\begin{equation}
\DRG F_R = - \gamma_F F_R,
\end{equation}
which in the component notation reads
\begin{equation}
\DRG F_{iR} = - \gamma_F^{ij} F_{jR}.
\end{equation}
Here, $\gamma_F$ is a matrix of anomalous dimension for a closed set of composite operators. 

The differential operator 
$\DRG$
is RG operator in the MS scheme and can be written in the
the following form
\begin{equation}
\DRG \equiv \tilde{\mathcal{D}}_{\mu} = \mathcal{D}_{\mu} + \beta_g \partial_g - \gamma_{\alpha} \partial_{\alpha},
\end{equation}
where $\mathcal{D}_{\mu} = \mu \partial_{\mu}$ and $\beta_g$ is beta function of coupling constant $g$. At the fixed point we have $g = g^{*}$, and the contribution of the beta function vanishes.
In time-coordinate representation, the equation of critical scaling for the closed set of renormalized operators $\{ F \}_R$ takes the form
\begin{equation}
\left[- \mathcal{D}_x + \Delta_t
\mathcal{D}_t \right] F_R = \Delta_F F_R,
\label{eq:critic_scaling}
\end{equation}
where $\Delta_t = - \Delta_{\omega}$ is critical dimension, and $\Delta_F$ is a matrix of critical dimensions. The critical dimension of a composite operator $\Delta_F$ can be obtained from the formula
\begin{equation}
\Delta_F = d_p [F] + \Delta_{\omega} d_{\omega} [F] + \gamma_F^* = d [F] - \gamma_{\alpha}^{*} d_{\omega} [F] + \gamma_F^*,
\end{equation}
where $\Delta_{\omega} = 2-\gamma_{\alpha}^{*} $, and anomalous dimensions $\gamma_F^{*}$ and $\gamma_{\alpha}^{*}$ are taken at the IR-stable fixed point $(g = g^*)$. The generalized model~A  exhibits a single with the coordinate (see \cite{Vasilev}, chapter 4.2)
\begin{equation}
g^{*} = \frac{3}{n+8} \varepsilon \bigg|_{n=2} = \frac{3}{10} \veps.
\end{equation}  
The matrix of anomalous dimension $\gamma_F$ is defined as
\begin{equation}
\gamma_F \equiv Z_F^{-1} \cdot (\widetilde{\mathcal{D}}_{\mu} Z_F),    \quad
\gamma_F = - \textrm{ UV-finite part} \left[ \widetilde{\mathcal{D}}_{\mu} Z^a \right],
\end{equation}
where the second relation is valid in the MS scheme \cite{Vasilev}. The relation can be further simplified in the leading order of perturbation theory 
\begin{equation}
\gamma_F = \veps g \partial_g Z^a = \varepsilon g \partial_g Q.
\label{eq:anom_dim}
\end{equation}

The diagonal matrix of the canonical dimensions $d_F$ is known 
(see Tab.~\ref{tab:can_dim_f}) and matrix $\gamma_F^{*}$ is calculated from the relation \eqref{eq:anom_dim} at the fixed point $g=g^{*}$. If the $\gamma_F^*$ was diagonal as well, the equation of critical scaling \eqref{eq:critic_scaling} would have the same structure and the diagonal elements of the matrix $\Delta_F$ would be interpreted as the critical dimensions of the corresponding operators $F_{iR}$. However, the blocks $\mathcal{Q}_i$ are non-zero square matrices, and the corresponding matrix of anomalous dimension $\gamma_F^{*}$ at the fixed points is therefore non-diagonal 
(and asymmetric). As a direct consequence, the operators $F_{iR}$ do not possess definite critical dimensions. Instead, we can form certain linear combinations of the latter and obtain
quantities with definite critical dimensions. Let us write
\begin{equation}
F'_{iR} = U_{ij} F_{jR},
\end{equation}
where $U$ is some matrix, which depends on parameters $\mu$ and $\alpha$. The substitution $F_R = U^{-1} F_R'$ in \eqref{eq:critic_scaling} yields an equation for new set of composite operators $F_{R}'$ with a new matrix \cite{Vasilev}
\begin{equation}
\Delta_F' = U \Delta_F U^{-1}, 
\end{equation}
where the final expression is derived for the massless model~A. The final value of diagonal element $\Delta_F'$ has been calculated numerically for each block separately
\begin{align}
\Delta_{F_1}' & = \{ \{ 8 - \frac{39}{5} \veps\}, \{ 8 - 4 \veps\},\{ 8 - 4 \veps \}, \{ 8 - 3 \veps\},\{ 8 - 3 \veps \}, \{ 8 - \frac{14}{5} \veps\}, \nonumber \\
& \{ 8 - \frac{12}{5} \veps\},\{ 8 - 2.653 \veps \},\{ 8 - 3.346 \veps \}, \{ 8 - 2.526 \veps \}, \{ 8 - 3.674 \veps \},  \nonumber \\ 
&  \{ 8 - 2.589 \veps \}, \{ 8 - 3.511 \veps \} \}, \label{eq:deltaF1}\\
\Delta_{F_2}' & = \{ \{ 8 - \frac{39}{5} \veps\}, \{ 8 - 4 \veps\},\{ 8 - 4 \veps \}, \{ 8 - 3 \veps\},\{ 8 - 3 \veps \}, \{ 8 - \frac{14}{5} \veps\}, \nonumber \\
& \{ 8 - \frac{12}{5} \veps \}, \{ 8 - 2.653 \veps \}, \{ 8 - 3.346 \veps \}, \{ 8 - 2.721 \veps \}, \{ 8 - 3.179 \veps \} \}
\label{eq:deltaF2} \\
\Delta_{F_3}' & = \{ \{ 8 - 4 \veps\}, \{ 8- \frac{12}{5} \veps \}, \{ 8- \frac{12}{5} \veps \}, \{ 8 - \frac{12}{5} \veps \}, \{ 8 - 3 \veps \},  \{ 8 - 3 \veps \},   \nonumber \\ 
& \{ 8 - 3 \veps \}, \{ 8 - \frac{14}{5} \veps \}, \{ 8 - \frac{64}{5} \veps \}, \{ 8 - \frac{11}{5} \veps \}, \{ 8 - \frac{29}{5} \veps \}, \{8 - \frac{31}{5} \veps \}, \nonumber \\
& \{ 8 - \frac{39}{5} \veps \}, \{ 8 - 3.527 \veps \}, \{ 8 - 3.269 \veps \}, \{ 8 - 2.583 \veps \}, \{ 8 - 2.154 \veps \} \} \label{eq:deltaF3} 
\end{align}
where the values are rounded to the third decimal place. The critical dimensions are calculated for the value $n=2$.
%
%
Let us note that the first seven values in the expressions \eqref{eq:deltaF1} and \eqref{eq:deltaF2}, are, in fact, exact, whereas in expression \eqref{eq:deltaF3} the first thirteen values are exact. Finally, our main interest is the behavior of viscosity. This can be related to the critical dimensions of source terms of corresponding composite operators. 
%
%

The critical dimension of the sources $\Delta_a'$ can be determined from the relation
\begin{equation}
\Delta_a' = d + \Delta_{\omega} - \Delta_F'.
\end{equation}
%
%
It is worthwhile to note, the analysis becomes simple for the case when a source is connected to the coefficient of viscosity. The presence of viscosity as a source reduces the set of operators (rows and columns in the mixing matrix) for the operators which are connected through the derivative with the same number of fields, e.g. $\partial_t (\ffi \ff) (\ff \ff) = \vf{10} +2 \vf{11} +\vf{12}  = 0$. The composite operators connected via derivative are linearly dependent and one of the operators can be excluded from the analysis. This property will provide a significant reduction in the calculation of the critical dimension for the viscosity coefficient. For completeness, we keep all operators in the work, and their form is reduced depending on the task. 
%
%

We have calculated the critical dimensions of a set of composite operators with the canonical dimension $8$ to the leading order. After verifying these results, the next task consists of a determination of the critical dimensions for operators with a lower canonical dimension and determining the behavior of the viscosity coefficient in the leading order of perturbation theory.

\section*{Conflict of Interest}
 The authors declare that they have no
conflicts of interest.

\subsection*{Acknowledgment}
The work was supported by VEGA grant No. 1/0535/21 of the Ministry of Education, Science, Research and Sport of the Slovak Republic and the Foundation for the Advancement of Theoretical Physics and Mathematics “BASIS” 19-1-1-35-1. 

\section*{Appendix}

Calculating the contributions of individual composite operators to the mixing matrix $\mathcal{Q}$ is a laborious process, so we will simplify this description to the present final form of the mixing matrix. More precisely, these are the parts $\mathcal{Q}_1$, $\mathcal{Q}_2$ and $\mathcal{Q}_3$ that contribute to the final values of the critical dimensions $\Delta_F'$.

In the calculation, we have employed standard quantum-field methods \cite{Vasilev,Zinn2002,Amit}. To simplify the notation,  we have
redefined the coupling constant in the calculations according to the prescription
\begin{equation}
   \frac{g}{(4\pi)^{d/2}} \longrightarrow g   .
\end{equation}

The part of mixing matrix (\ref{eq:qmatrix}), which is denoted as $\mathcal{Q}_1$ has following form

\begin{eqnarray*}
\fontsize{12}{10}\selectfont
\left(
\begin{smallmatrix}
 \frac{n+6}{3} & \frac{8}{3} & \frac{n+2}{8 } & 0 & \frac{n+2}{4 } & \frac{n+2}{8 } & 0 & 0 & 0 & \frac{n+2}{4 \alpha  } & 0 & \frac{n+2}{4 \alpha} & 0 \\[2mm]
 1 & \frac{2 (n+5)}{3} & \frac{3}{8 } & \frac{1}{2 } & \frac{1}{4 } & \frac{1}{8 } & \frac{1}{4 } & 0 & 0 & \frac{1}{4 \alpha } & \frac{1}{2 \alpha  } & \frac{3}{4 \alpha  } & 0 \\[2mm]
 0 & 0 & \frac{5 (n+8)}{12 } & \frac{2}{3 } & \frac{n+4}{6 } & \frac{n+4}{12 } & \frac{1}{3 } & 0 & 0 & -\frac{n+4}{6 \alpha  } & -\frac{2}{3 \alpha } & -\frac{n+8}{6 \alpha  } & 0 \\[2mm]
 0 & 0 & \frac{1}{2 } & \frac{7 n+44}{12 } & \frac{5}{4 } & \frac{1}{6 } & \frac{1}{3 } & \frac{1}{4 } & \frac{n+4}{4 } & \frac{1}{6 \alpha } & \frac{1}{3 \alpha  } & \frac{1}{2 \alpha  } & 0 \\[2mm]
 0 & 0 & \frac{n+2}{6 } & \frac{7}{3 } & \frac{4 n+15}{6 } & \frac{n+2}{6 } & 0 & \frac{1}{2 } & 1 & \frac{n+2}{6 \alpha  } & 0 & \frac{n+2}{6 \alpha  } & 0 \\[2mm]
 -\frac{8}{9 } & -\frac{16}{9 } & -\frac{n+2}{12 } & 0 & -\frac{n+2}{6 } & \frac{3 n+14}{12 } & \frac{4}{3 } & 0 & 0 & -\frac{n-2}{6 \alpha  } & \frac{4}{3 \alpha  } & -\frac{n+2}{6 \alpha  } & 0 \\[2mm]
 -\frac{4}{9 } & -\frac{4 (n+4)}{9 } & -\frac{1}{4 } & -\frac{1}{3 } & -\frac{1}{6 } & \frac{7}{12 } & \frac{2 n+11}{6 } & 0 & 0 & \frac{1}{6 \alpha  } & \frac{n+3}{3 \alpha  } & -\frac{1}{2 \alpha  } & 0 \\[2mm]
 0 & \frac{4 (n+2)}{9 } & 0 & \frac{1}{3 } & \frac{1}{6 } & 0 & \frac{n+2}{3 } & \frac{4 n+13}{6 } & \frac{5}{3 } & 0 & -\frac{n+2}{3 \alpha  } & 0 & 0 \\[2mm]
 \frac{4}{9 } & \frac{8}{9 } & 0 & \frac{n+4}{12} & \frac{1}{12 } & \frac{1}{3 } & \frac{2}{3 } & \frac{13}{12 } & \frac{5 n+36}{12 } & -\frac{1}{3 \alpha} & -\frac{2}{3 \alpha  } & 0 & 0 \\[2mm]
 0 & 0 & \frac{\alpha  (n+2)}{12 } & 0 & \frac{\alpha  (n+2)}{6 } & \frac{\alpha  (n+2)}{12 } & 0 & 0 & 0 & \frac{3 n+14}{6 } & \frac{8}{3 } & \frac{n+2}{6 } & 0 \\[2mm]
 0 & 0 & \frac{\alpha }{4 } & \frac{\alpha }{3 } & \frac{\alpha }{6 } & \frac{\alpha }{12 } & \frac{\alpha }{6 } & 0 & 0 & \frac{7}{6 } & \frac{2 n+11}{3 } & \frac{1}{2 } & 0 \\[2mm]
 0 & 0 & -\frac{\alpha  (n+8)}{12 } & -\frac{2 \alpha }{3 } & -\frac{\alpha  (n+4)}{6 } & -\frac{\alpha  (n+4)}{12 } & -\frac{\alpha }{3 } & 0 & 0 & \frac{n+4}{6 } & \frac{2}{3 } & \frac{n+8}{2 } & 0 \\[2mm]
 0 & 0 & 0 & 0 & 0 & 0 & 0 & 0 & 0 & 0 & 0 & 0 & n+14 \\[2mm]
\end{smallmatrix}
\right)
\end{eqnarray*}

The part of mixing matrix (\ref{eq:qmatrix}), which is denoted as $\mathcal{Q}_2$ has form 
\begin{eqnarray*}
\fontsize{12}{10}\selectfont
\left(
\begin{smallmatrix}
 n+14 & 0 & 0 & 0 & 0 & 0 & 0 & 0 & 0 & 0 & 0 \\[2mm]
 0 & \frac{n+8}{2 } & \frac{n+4}{6 } & \frac{2}{3 } & \frac{\alpha  (n+8)}{12 } & \frac{\alpha  (n+4)}{12 } & \frac{\alpha }{3 } & 0 & 0 & \frac{\alpha (n+4)}{6 } & \frac{2 \alpha }{3 } \\[2mm]
 0 & \frac{n+2}{6 } & \frac{3 n+14}{6 } & \frac{8}{3 } & -\frac{\alpha  (n+2)}{12 } & -\frac{\alpha  (n+2)}{12 } & 0 & 0 & 0 & -\frac{\alpha  (n+2)}{6 } & 0 \\[2mm]
 0 & \frac{1}{2 } & \frac{7}{6 } & \frac{2 n+11}{3 } & -\frac{\alpha }{4 } & -\frac{\alpha }{12 } & -\frac{\alpha }{6 } & 0 & 0 & -\frac{\alpha }{6} & -\frac{\alpha }{3 } \\[2mm]
 0 & \frac{n+8}{6 \alpha  } & \frac{n+4}{6 \alpha  } & \frac{2}{3 \alpha  } & \frac{n+8}{4 } & -\frac{n+4}{12 } & -\frac{1}{3 } & 0 & 0 & -\frac{n+4}{6} & -\frac{2}{3 } \\[2mm]
 0 & \frac{n+2}{6 \alpha  } & \frac{n+6}{6 \alpha  } & \frac{4}{3 \alpha  } & \frac{n+2}{12 } & \frac{5 n+18}{12 } & \frac{4}{3 } & 0 & 0 & \frac{n+2}{6} & 0 \\[2mm]
 0 & \frac{1}{2 \alpha} & \frac{1}{2 \alpha} & \frac{n+5}{3 \alpha} & \frac{1}{4} & \frac{3}{4} & \frac{2 n+13}{6 } & 0 & 0 & \frac{1}{6 } & \frac{1}{3 } \\[2mm]
 0 & 0 & 0 & -\frac{n+2}{3 \alpha  } & 0 & 0 & \frac{n+2}{3} & \frac{4 n+15}{6} & \frac{7}{3} & \frac{1}{2} & 1 \\[2mm]
 0 & 0 & -\frac{1}{3 \alpha  } & -\frac{2}{3 \alpha} & 0 & \frac{1}{3 } & \frac{2}{3 } & \frac{5}{4 } & \frac{7 n+44}{12 } & \frac{1}{4 } & \frac{n+4}{4 } \\[2mm]
 0 & -\frac{n+2}{6 \alpha  } & -\frac{n+2}{6 \alpha  } & 0 & \frac{n+2}{6 } & \frac{n+2}{6 } & 0 & \frac{1}{6 } & \frac{1}{3 } & \frac{4 n+13}{6} & \frac{5}{3 } \\[2mm]
 0 & -\frac{1}{2 \alpha  } & -\frac{1}{6 \alpha  } & -\frac{1}{3 \alpha  } & \frac{1}{2 } & \frac{1}{6 } & \frac{1}{3 } & \frac{1}{12 } & \frac{n+4}{12 } & \frac{13}{12 } & \frac{5 n+36}{12 } \\[2mm]
\end{smallmatrix}
\right)
\end{eqnarray*}

The part of mixing matrix (\ref{eq:qmatrix}), which is denoted as $\mathcal{Q}_3$ has form 

\begin{eqnarray*}
\fontsize{10}{8}\selectfont
\left(
\begin{smallmatrix}
 \frac{4n+80}{3} & 0 & 0 & 0 & 0 & 0 & 0 & 0 & 0 & 0 & 0 & 0 & 0 & 0 & 0 & 0 & 0 \\[2mm]
 0 & \frac{2 n + 28}{3} & 0 & 0 & 0 & 0 & 0 & 0 & 0 & 0 & 0 & 0 & 0 & 0 & 0 & 0 & 0 \\[2mm]
 0 & \frac{n+2}{3} & n +\frac{26}{3} & \frac{16}{3} & 0 & 0 & 0 & 0 & 0 & 0 & 0 & 0 & 0 & 0 & 0 & 0 & 0 \\[2mm]
 0 & 1 & \frac{4}{3} & n +\frac{38}{3} & 0 & 0 & 0 & 0 & 0 & 0 & 0 & 0 & 0 & 0 & 0 & 0 & 0 \\[2mm]
 0 & 0 & 0 & 0 & \frac{n+8}{3 } & 0 & 0 & 0 & 0 & 0 & 0 & 0 & 0 & 0 & 0 & 0 & 0 \\[2mm]
 0 & 0 & 0 & 0 & 0 & \frac{n+2}{3 } & 0 & 0 & 0 & 0 & 0 & 0 & 0 & 0 & 0 & 0 & 0 \\[2mm]
 0 & 0 & 0 & 0 & 0 & \frac{1}{3 } & \frac{2}{3 } & 0 & 0 & 0 & 0 & 0 & 0 & 0 & 0 & 0 & 0 \\[2mm]
 0 & 0 & 0 & 0 & 0 & 0 & 0 & \frac{n+4}{3 } & \frac{4}{3} & 0 & 0 & 0 & 0 & 0 & 0 & 0 & 0 \\[2mm]
 0 & 0 & 0 & 0 & 0 & 0 & 0 & \frac{2}{3 } & \frac{n+6}{3} & 0 & 0 & 0 & 0 & 0 & 0 & 0 & 0 \\[2mm]
 0 & 0 & 0 & 0 & 0 & 0 & \frac{n+2}{3 } & 0 & 0 & \frac{n+4}{3} & \frac{4}{3} & 0 & 0 & 0 & 0 & 0 & 0 \\[2mm]
 0 & 0 & 0 & 0 & 0 & \frac{1}{3} & \frac{2}{3} & 0 & 0 & \frac{2}{3} & \frac{n+6}{3} & 0 & 0 & 0 & 0 & 0 & 0 \\[2mm]
 0 & 0 & 0 & 0 & 0 & 0 & 0 & 0 & 0 & \frac{2 n+8}{3} & \frac{8}{3} & \frac{2 n+8}{3} & 0 & 0 & 0 & 0 & \frac{8}{3} \\[2mm]
 0 & 0 & 0 & 0 & \frac{n+2}{6} & \frac{3 n+2}{18} & -\frac{4}{9} & \frac{2 n+4}{3} & 0 & -\frac{2}{9 } & -\frac{4}{9 } & 0 & \frac{6 n+20}{9} & \frac{16}{9} & \frac{8}{9} & \frac{16}{9} & 0 \\[2mm]
 0 & 0 & 0 & 0 & \frac{1}{2} & \frac{1}{18 } & \frac{-n-1}{9} & \frac{2}{3 } & \frac{4}{3 } & \frac{-n-3}{9} & -\frac{2}{9} & 0 & \frac{10}{9} & \frac{4n+28}{9 } & \frac{4n+12}{9} & \frac{8}{9 } & 0 \\[2mm]
 0 & 0 & 0 & 0 & 0 & 0 & \frac{n+2}{18 } & \frac{1}{3 } & \frac{2}{3} & \frac{1}{3} & \frac{2}{3} & \frac{-n-2}{18 } & 0 & \frac{n+2}{9} & \frac{n+6}{3} & \frac{8}{3} & \frac{2n+4}{9} \\[2mm]
 0 & 0 & 0 & 0 & 0 & \frac{1}{18 } & \frac{1}{9 } & \frac{1}{6} & \frac{n+4}{6} & \frac{1}{6} & \frac{n+4}{6} & \frac{1}{6} & \frac{1}{9} & \frac{2}{9} & 1 & \frac{2n+10}{3} & \frac{1}{3} \\[2mm]
 0 & 0 & 0 & 0 & 0 & 0 & 0 & 0 & 0 & \frac{n+4}{9 } & \frac{4}{9} & \frac{n+8}{9 } & 0 & 0 & \frac{2 n+8}{9 } & \frac{8}{9} & \frac{2n+16}{9 } \\[2mm]
\end{smallmatrix}
\right)
\end{eqnarray*}



\end{document}